\documentclass[preprint2]{aastex}

\slugcomment{To be submitted to ApJ}

\shorttitle{Modelling Mira's wind}
\shortauthors{Ryde et al.}

\begin{document}

\title{Modelling CO emission from Mira's wind}

\author{N. Ryde}
\affil{Uppsala Astronomical Observatory, Box 515, SE-751 20, Uppsala, Sweden}
\email{ryde@astro.uu.se}

\and

\author{F.L. Sch\"oier}
\affil{Stockholm Observatory, SE-133 36 Saltsj\"obaden, Sweden}
\email{fredrik@astro.su.se}

\begin{abstract}

We have modelled the circumstellar envelope of {\it o} Ceti (Mira) using new 
observational constraints. These are obtained from 
photospheric light scattered in near-IR 
vibrational-rotational lines of circumstellar CO molecules at 
$4.6\,\mbox{$\mu$m}$: absolute fluxes,
the radial dependence of the scattered intensity, and two line ratios.
Further observational constraints are provided by ISO
observations of far-IR emission lines from highly excited rotational states 
of the ground vibrational state of CO, and radio observations of lines from 
rotational levels of low excitation of CO.
A code based on  the Monte-Carlo technique
is used to model the circumstellar line emission. 

The vibrational-rotational lines are sensitive to the radiation field, 
whereas the pure rotational lines, such as the rotational lines of low 
excitation  measured 
at radio wavelengths and the rotational lines from highly excited states 
observed with ISO, 
are usually more sensitive to the temperature structure. 
These rotational lines 
have been the prime probe in most earlier investigations.

We find that it is  possible to model the radio and ISO fluxes,   
as well as the highly asymmetric radio-line profiles, reasonably well with 
a spherically symmetric and smooth stellar wind model. 
However, it is  not possible to reproduce the observed NIR line fluxes 
consistently with a `standard model' of the stellar wind.  
This is probably due to incorrectly specified  conditions of the inner 
regions of the wind 
model, since the stellar flux needs to be larger than what is 
obtained from the standard model
at the point of scattering, i.e., 
the intermediate regions at approximately 100$-$400\,R$_*$ 
(2$\arcsec$$-$7$\arcsec$) away from the star. 
Thus, the optical depth in the vibrational-rotational lines  
from the star to the point of scattering has to be 
decreased. This can be accomplished in several ways. For instance, the gas 
close to the star (within approximately 2$\arcsec$) could be in such a 
form that light is able to pass through, either due to the medium being 
clumpy or by the matter being in radial structures 
(which, further out, developes into more smooth or shell-like 
structures). 
Further observations of the gas in the stellar wind close to Mira are required 
to resolve this problem.

The model circumstellar envelope, which  reproduces the 
observables reasonably well, has a mass-loss rate of 
2.5$\times$10$^{-7}$\,M$_\odot$\,yr$^{-1}$, and a turbulent 
velocity of 1.5\,km\,s$^{-1}$, given a terminal expansion velocity of the wind 
of 2.5\,km\,s$^{-1}$. 
\end{abstract}


\section {INTRODUCTION\label{oceti}}

The inner regions of circumstellar envelopes around asymptotic giant branch 
stars are still enigmatic. These regions, located further out than the 
pulsating
outer atmosphere but typically within 100 stellar radii, are of interest 
for several reasons; 
The circumstellar wind is accelerated here to its terminal velocity, the 
chemistry is `frozen out' approximately at its photospheric values, and
dust nucleation
and growth take place. This is important for the mass loss of the star,
a process we do not 
understand how and why it occurs. For example, we do not know whether matter
is ejected in puffs or in a steady, homogeneous wind. The mass loss returns
matter to the interstellar medium and plays a vital role in the cosmic 
cycling and enrichment of matter, see for example \citet{IAU177}. 
In order to study these inner regions we need to observe them in detail.
However, very few observational constraints exist on
the conditions in these complex regions. 

By combining 
observations, sampling gas located at a distance on the sky of 
approximately 1$\arcsec$ and beyond, and a detailed modelling of the 
circumstellar envelopes one might be able to put constraints on the inner 
regions.
\citet{ryde:apj} observed scattered photospheric light in circumstellar
vibration-rotational lines of CO 2$\arcsec$$-$7$\arcsec$ away from Mira.
They analysed the emission lines 
from the circumstellar 
envelope using a simple analytic approach, 
an analysis which suggested a time constant
mass-loss rate, when averaged over 100 years, over the past 1200 years.
In this paper we present a detailed analysis both of 
these observations and of those which tell us about the properties 
of the whole envelope.
The modelling 
presented here provides
a better insight into the conditions close to the star, and gives a larger
degree of flexibility to test different models.

{\it o} Ceti (Mira or HD 14386) is the prototype of the mira class of 
long-period variables with characteristic pseudo-periodic brightness 
variations in time. They are cool, red giants on the asymptotic giant branch 
(AGB) with considerable mass-loss rates and circumstellar envelopes (CSEs) of 
large extension. The visual-light variations are related to large-amplitude 
radial pulsations which cause a large variation of the effective temperature. 
Note, that 
{\it o} Ceti is a binary system.
The angular distance between Mira~A and the hot, compact companion star
Mira~B (VZ~Ceti) is 0.6$\arcsec$ (Karovska et al. 
1993; 1997)\nocite{karovska:93}\nocite{karovska:97}.
The companion could be a white dwarf with a mass of about 1\,M$_\odot$, 
a luminosity of 2\,L$_\odot$, and a temperature of more than 
30000\,K [these parameters are based on an analysis of International Ultraviolet 
Explorer (IUE) spectra by 
\citet{stickland}]. \citet{danchi:94} suggest it is embedded in an 
accretion disk, giving rise to 
an abnormal illumination of Mira~A.

In general, the kinematics, the density and the temperature structures of 
CSEs of AGB stars seem to be relatively simple, with an over-all spherical 
symmetry prevailing (see, e.g., Olofsson 1996\nocite{ho:rev}). The stable and
abundant CO molecule is self-shielding \citep{glassgold} and therefore 
of approximately constant 
abundance trough out the envelope.  
However, the chemical structure is certainly quite complex,
restricting us to schematic modelling of these objects [see for example 
\citet{glassgold}].
The lack of spatial resolution in the observations, nonetheless suggesting 
more complex 
geometries (e.g., Stanek et al. 1995\nocite{stanek:95}), also hampers the 
analysis. 

The observations of CO lines at millimeter wavelength in Mira's wind indicate 
some complexity. An asymmetry is clearly visible in the radio-line profiles. 
A number of suggestions in the literature have been made for multi-wind 
scenarios, such as
winds of different characteristics, jets, or other phenomena 
which introduce dramatic variations in the envelope structure with distance.
%
Thus, several combinations exist of mass-loss rates, expansion and 
turbulent velocities, which are able to reproduce the radio-line profiles 
fairly well. 
This means, for instance, that there is no consensus as regards the actual 
expansion velocity of Mira's wind. 
\citet{crosas:apss} experiment with an expansion velocity ($v_\mathrm{e}$) of 
approximately 2\,km\,s$^{-1}$ 
and a turbulent velocity ($v_\mathrm{t}$) of approximately 
4\,km\,s$^{-1}$ in their version of the inner wind, while 
\citet{young} arrived at $v_\mathrm{e}$$=$4.8\,km\,s$^{-1}$. However, 
\citet{young} points out that the wings show an expansion velocity of 
10\,km\,s$^{-1}$. \citet{knapp:98} model a fast outer wind with a mass-loss 
rate of $\dot{M}$$=$4.4$\times$10$^{-7}$\,M$_\odot$\,yr$^{-1}$, and an 
expansion velocity of 6.7$\pm$1.0\,km\,s$^{-1}$. An inner wind is supposed to 
be a resumed wind in analogy with the detached shells found in four carbon 
stars \citep{ho:96a}. 
This slow wind component has a lower mass-loss rate and a much lower expansion 
velocity than what is found normally in Miras;  
$\dot{M}$$=$9.4$\times$10$^{-8}$\,M$_\odot$\,yr$^{-1}$ and 
$v_\mathrm{e}$$=$2.4$\pm$0.4\,km\,s$^{-1}$.
Planesas et al. (1990a) \nocite{planesas:I} interpret the 
CO $J$$=$1$\rightarrow$0 and $J$$=$2$\rightarrow$1 lines as 
partly originating from a spherically
symmetric CSE with $v_\mathrm{e}$$=$3\,km\,s$^{-1}$. 

Numerous high-resolution, optical and infrared wavelength measurements  
provide evidence that Mira itself is elongated
(e.g., Karovska et al. 1991, 1997\nocite{karovska:97}\nocite{karovska:91};
Haniff et al.\ 1992\nocite{haniff:92};
Wilson et al.\ 1992\nocite{wilson:92};
Quirrenbach et al.\ 1992\nocite{quirren:92}).
\citet{karovska:97} suggest that the asymmetry could be due to unresolved 
bright spots or non-radial pulsation, or even a wind interaction with Mira B. 
As an alternative to an asymmetric atmosphere, 
\citet{danchi:94} discussed the possibility of the observed asymmetries being caused by non-uniform 
dust shells close to the star. 
Also Planesas et al. (1990a,b)\nocite{planesas:I,planesas:II} interpreted 
their asymmetric CO $J$$=$1$\rightarrow$0 and $J$$=$2$\rightarrow$1
line profiles and brightness maps, which are at a spatial resolution of 
6$\arcsec$, as originating partly from a very slow, partially collimated wind, 
due to high densities of gas in the equatorial plane close to the star. 

We will, nevertheless, assume a spherically symmetric model for the CSE since 
the situation is unclear as regards the physical origin of the complexity, 
and its dependence on the distance from the star. 
Also, \citet{ryde:apj} concluded from their near-IR (NIR) CO line data 
that they do not necessarily
indicate a non-symmetric, inner wind.   

\section{OBSERVATIONAL DATA}

Our modelling of Mira's wind will be compared with three sets of 
observations probing different regions of the CSE.

{\it First}, spatially and spectrally resolved, vibrational-rotational line 
emission from  the CO fundamental band lying at 4.6\,$\mu$m is used as a 
probe of the inner regions \citep{ryde:apj}. The lines observed are R(1), R(2), 
and R(3). These emission lines consist of 
photospheric light scattered  by CO molecules in the intermediate regions
(approximately 100$-$400\,R$_*$) of the CSE around Mira.
These observations are the most important ones in the discussion presented 
here, providing a new constraint to the modelling of the CSE.
The emission lines were detected in observations with the Phoenix spectrometer 
mounted on the 4\,m Mayall telescope at Kitt Peak. The reduction and absolute
flux calibration of these observations are described by \citet{ryde:apj}.
The uncertainties in the absolute fluxes are probably lower than $\pm 50\%$, which we
regard as a conservative estimate. 

{\it Second}, as a further constraint on our model we have retrieved 
far-IR (FIR) observations of Mira, which probe regions within 2$\arcsec$ 
({\small
\raisebox{-0.02cm}{\begin{minipage}{0.2cm} 
\raisebox{-0.2cm}{$< $} \\ 
\raisebox{0.08cm}{$\sim$ }
\end{minipage}} }
100\,R$_*$) of the star, 
obtained with the Infrared Space Observatory (ISO)
(Kessler et al. 1996\nocite{kessler}).
These high-lying rotational lines of $^{12}$CO
are retrieved from the ISO-data archive\footnote{
This project is partly based on observations with ISO, an ESA 
project with instruments funded by ESA Member States (especially the 
PI countries: France, Germany, the Netherlands and the United Kingdom) 
and with the participation of ISAS and NASA. http://isowww.estec.esa.nl/}.   
The Long Wavelength Spectrometer (LWS, Clegg et al. 1996\nocite{LWS}) was used 
in the grating mode (LWS01) providing a mean spectral
resolution element of approximately 0.7\,$\mu$m. We will discuss only the 
line-integrated fluxes. 
The 32 minutes of observations were performed in the morning
the 4$^{\mathrm {th}}$ July 1997 during ISO revolution 596.  
The reductions were made using the most recent pipeline, basic reduction 
package OLP (v.7) and the ISO Spectral Analysis Package (ISAP v.1.5).  
The pipeline processing of the data, such as the flux calibration, 
is described by \citet{swinyard}; the combined absolute and systematic 
uncertainties in the fluxes are large, of the order $\pm$50\%. 
Mira is considered to be a faint source in the ISO-LWS reduction procedures
(Flux$<50\,\mbox{Jy}$ at 
$120\,\mbox{$\mu$m}$). The observations consist of 
weak emission lines on a strong continuum. A large uncertainty
is introduced in the subtraction of the continuum.
We have measured the emission intensities in the same manner as 
is described in  \citet{FL} for IRAS\,15194-5115. The lines studied are $J$$=$19$\rightarrow$18, 
$J$$=$18$\rightarrow$17, $J$$=$17$\rightarrow$16, $J$$=$16$\rightarrow$15,
at 137.2, 144.7, 153.3, and 162.9\,$\mu$m, respectively. 
The line-integrated fluxes measured 
yield a mean flux of approximately $(2\pm 1)\times$10$^{-20}$\,W\,cm$^{-2}$
for the four lines. 

{\it Third}, radio observations, made with the James Clerk Maxwell Telescope 
(JCMT) on Hawaii, of the 
CO  $J$$=$2$\rightarrow$1, $J$$=$3$\rightarrow$2, and 
$J$$=$4$\rightarrow$3 lines are retrieved from 
the public archive of the JCMT\footnote{http://www.jach.hawaii.edu/}. 
These data are taken at face value. By comparing different observations we 
estimate the uncertainty in the intensity scale to be 
approximately $\pm20\%$. The intensity scale is given in main-beam brightness
temperature, $T_{\mathrm{mb}}$.
These data provide constraints on the outer regions of the CSE, i.e., 
approximately 1000\,R$_*$ away from the star and beyond.

In Tables \ref{resultat1}, and \ref{resultat2} the 
observational results of the NIR, ISO, and JCMT observations are presented. 
In Table\,\ref{resultat1}, 
the mean values of the NIR line intensities obtained west, east, north, and south 
\citep{ryde:apj}
are given, as well as the R(1)/R(3) and 
R(2)/R(3)-line ratios. The intensities are means for a position 
(2$\pm$0.5)$\arcsec$ away from the star, the ratios are means over the distance 
2$-$3.4$\arcsec$. Table\,\ref{resultat1} also presents the radial brightness 
distribution of the NIR line emission given as the exponent of the 
intensity power-law (d$\log$ I/d$\log \beta$) of the 
R(2) vibrational-rotational line and is a mean over the entire range
($\beta$ is the angular distance from the star on the sky). 
The  uncertainties quoted for the line ratios and the slope are pure measurement uncertainties. The R(1)/R(3) 
ratio in Table\,\ref{resultat1} may be uncertain due to systematic 
uncertainties in R(1), cf. \citet{ryde:apj}. The ISO flux in Table\,\ref{resultat2} 
is given as the mean of the four measured lines in order to reduce the uncertainty
since the data is noisy and the line intensities should not vary much.


\section{THE MODEL OF THE CIRCUMSTELLAR ENVELOPE\label{modell}}

We have used a Monte-Carlo (MC) approach to the multi-level, non-LTE, 
radiative-transfer problem when modelling the circumstellar wind of Mira.
We have applied the methodology of \citet{bernes} to the transfer of 
line-radiation of CO in a spherically symmetric and smooth CSE, formed by a 
constant mass-loss rate, and expanding  at a constant velocity. 
The inner and outer radii of the CSE
are used as the input for the modelling. For a more detailed description of the 
modelling, see \citet{PhD}. 
 
The model is constructed by iteratively solving the equations of 
radiative transfer and the statistical equilibrium equations
for the level populations of $^{12}$CO. The solution is found when
a steady state of the populations, as a function of radial distance from 
the star, is achieved. The source  function of the entire wind is thereby 
known, which makes possible a straight-forward calculation of the spectrum
by the solution of the radiative transfer through the envelope along the line-of-sight,
resulting in predicted intensities of the observed spectral lines as a 
function of the distance from the star. 


A draw-back of Monte-Carlo simulations in general is the slow convergence, 
typically scaling with $\sqrt{\mathrm{N_{iter}}}$. On the other hand, the 
advantages are several. 
The programming can be done in close relation to the physical reality; 
the radiation field can be taken into account in a correct
manner, and inhomogeneities and complex geometries are relatively easily 
implemented.
Velocity fields and gradients of any size are allowed in the programming.  

In our MC model, complete angular and frequency redistribution 
is assumed, the absorption and the emission profiles being the same. 
Thus, the local emission is assumed to be isotropic and the scattering 
is assumed to be incoherent.
The distribution of frequency shifts given
by the emission profile is defined by a gaussian Doppler profile:

\begin{equation}
\label{doppler}
\Phi(\nu)=\frac{1}{\sigma \sqrt{\pi}} \exp \left[\frac{-(\nu-\nu_{\mathrm{0}}-\vec{v}\cdot \vec{n} \, \nu_{\mathrm{0}}/c)^2}{\sigma^2}\right].
\end{equation}
The width of this profile, $\sigma$, is given by the local, thermal 
broadening and the
micro-turbulent broadening. 
The velocity field in the expanding wind will allow 
shifts from the line centre
as given in the laboratory frame, $\nu_\mathrm{0}$. The velocity field is represented by the vector
$\vec{v}$ and the direction of the beam, i.e. the random direction of one emitted photon, 
by the unit vector $\vec{n}$. 

For the rotational 
transitions, with spontaneous de-excitation rates of the order of 
$10^{-7}$\,s$^{-1}$, the assumption of complete frequency redistribution (CRD)
is
a good assumption, since the collisions will have time to reset the 
frequency-memory of the excited atoms.
However, for the vibrational-rotational transitions, with rates of the order of 
$10$\,s$^{-1}$, this assumption may be a fallacy. For these infrared transitions, 
photons are more likely coherently scattered in frequency. 
The more scattering events, i.e., the higher the radial optical depth in the 
wind, the more important it
is that the model is made to properly take into account coherent scattering in the model.

By dividing every NIR transition into a large number of subdivisions, 
but conserving the over-all line profile, 
we are able to simulate coherent scattering in the MC code. The absorption and 
the emission profiles are assumed to be the same. We find that for optical 
depths of the order of 50 or less, the difference between a CRD and a 
coherent description of the NIR transitions is less than the numerical noise, 
which is intrinsic to the MC method. For a Mira model with a CSE with an inner radius 
at 13\,R$_*$, the radial optical depth is approximately 40, which motivates 
our approach to run the model with all transitions in CRD. 

\section{MODEL PARAMETERS}

Measurements of the multi-wavelength angular diameter of Mira
\citep{haniff:95}, combined with trigonometric
parallaxes measured with the Hipparcos satellite \citep{leeuwen:97}, 
suggest a mean radius of (464$\pm$80)\,R$_\odot$.
The distance to Mira is ($128^{+20}_{-15}$)\,pc \citep{hipp}.
>From \citet{mahler:97} one can derive a bolometric luminosity 
$L_{\mathrm{tot}}$$=$8900\,L$_\odot$, a temperature  
$T_\mathrm{eff}$$=$2400\,K and a radius of 
550\,R$_\odot$$=$3.8$\times$10$^{13}$\,cm for the phase of {\it o} Ceti at 
October 29, 1998, the date the NIR observations were made, see Figure \ref{fig3}. 
Mira was at its maximum radius at this epoque according 
to \citet{mahler:97}.
The ISO and radio data were observed at other times, but are not as 
sensitive to the radiation field, and thereby to the phases of Mira, as the NIR 
data. These values of the model parameters of Mira are consistent with the 
corresponding values obtained by, among others,
\citet{danchi:94} and \citet{haniff:95}. These authors arrive at similar 
temperatures; a mean (over phases) effective temperature of 
$T_{\rm{eff}}$$\sim$2800\,K. 
The terminal expansion velocity is set to  be $v_\mathrm{e}$$=$2.5\,km\,s$^{-1}$ 
\citep{knapp:98}.
The model parameters we have adopted are given in Table\,\ref{parameters}.

The radiative energy input to the wind system originates from the central star 
and is specified by a luminosity and an effective temperature of the star
and  by a luminosity and effective temperature of the surrounding dust, 
cf. Table\,\ref{parameters}. Since the dust is concentrated close to the star
\citep{bester:91}, the ambient radiation from the dust further out in the 
CSE is considered to be small and it is not taken into account in our model.
The dust will absorb part of the stellar flux and re-radiate it in the 
infrared as thermal radiation.
\citet{lopez:97} 
fit their observations with two spherically symmetric dust shells at 3 and 10 
stellar radii, with inhomogeneities or clumps embedded in the 
spherically symmetric dust shells.
\citet{danchi:94} 
fit their observations with a dust distribution
which peaks 
at 2.7 stellar radii away from the star, and where the temperature is 
1050$-$1300\,K, depending on phase.
Also \citet{bester:91} deduced 
an  inner extension of the dust shell of about three stellar radii, 
and a dust temperature of 1200\,K.
Thus, the spatial dust distribution peaks at approximately 3 stellar
radii with a temperature of about 1100\,K, leading to an upper limit of the 
re-radiated luminosity of 30\% of the total bolometric luminosity; 
2700\,L$_\odot$. \citet{haniff:95} found, however, that the circumstellar 
dust in Mira contributes less than 20\% of the bolometric luminosity. 
We therefore set the dust luminosity to be 1800\,L$_\odot$, implying a
stellar contribution of 7100\,L$_\odot$.
The former radiation, which is mainly in the infrared, is important for 
pumping vibrational states, leading to non-thermal level populations of the 
rotational states which could affect the NIR line intensities.
Hence, we model the inner energy source with a spectral energy distribution 
consisting of two black-bodies, using the temperatures and luminosities above, 
see also Table\,\ref{parameters}.




The kinetic temperature structure of the gas affects the CO level populations 
through collisions, modelled through collisional rates [see the discussion by \citet{FL}] which
are functions of the temperature. The temperature also affects the thermal 
broadening of the absorption and emission profiles. 
These are, however, mostly determined by the larger turbulent velocity.
The gas temperature structure of the CSE 
can be calculated in a self-consistent way in our model, by solving the 
energy balance equation of the system
(cf. Ryde et al. 1999c\nocite{FL}).
However, in the present discussion we select,  for simplicity,  
a structure of the gas kinetic temperature in the wind according to:
\begin{eqnarray}
\label{T}
T(r) & = & T_0 \left(\frac{10^{15}\,\mbox{cm}}{r}\right)^\zeta,
\end{eqnarray} 
where the exponent $\zeta$$\approx$1, see for example \citet{kahane:94}. 
$T_0$ is the temperature at 10$^{15}$\,cm and $r$ is the distance 
from the star in centimeters. This kinetic temperature structure and the 
excitation temperature structures calculated by the model are shown in 
Figure\,\ref{model2}b.

Mira~B certainly also adds mechanical energy to the circumstellar environment,
which will lead to an additional heating. The NIR lines are, however, 
radiatively excited and not greatly affected by the kinetic temperature of the 
surrounding gas/dust. The radio lines may, however, in principle be affected 
by their kinetic environment, but are formed further out where the  
influence of the binary star can be considered to be small.
 
The CSE is characterised by the mass-loss rate of the star and the  
expansion velocity of the wind, which together give the density as a function 
of the radial distance from the star. The density of CO will depend on the 
fractional abundance of CO relative to the abundance of hydrogen molecules, 
H$_2$. 
This fractional abundance of $^{12}$CO molecules is assumed to be
$f_\mathrm{CO}$$=$5$\times$10$^{-4}$ \citep{knapp:98} throughout the wind. 
This is a value generally used for oxygen-rich envelopes, and the same value 
was adopted in  \citet{ryde:apj}. 
In the case of Mira the mass-loss rate is relatively small, leading to an 
increased importance of the radiation field, especially for the NIR lines; 
see the discussion on the dusty envelope of the object 
IRAS\,15194-5115 with a high mass-loss rate, 
where the vibrational-rotational transitions are found
to be unimportant \citep{FL}.

The $^{12}$CO molecule data are the same as those used by \citet{FL}.
We use 30 rotational levels for each of the ground and first
vibrational states. This is generally sufficient in order to treat the CO 
excitation in CSEs properly, cf. \citet{FL}.
The collisional rates of the NIR vibrational-rotational transitions are
much smaller than the radiative de-excitation rates and they are set at zero 
in the model (see for example Ryde et al. 1999a\nocite{ryde_CO2}). 
The collisional 
rates of pure rotational transitions are non-zero. 
\citet{kirby} provide oscillator strengths, which for our transitions are consistent with
\citet{hure}\footnote{Note the missing cube in their Eq. 3}.



The photo-dissociation radius of the CO envelope ($r_\mathrm{p}$) 
is determined by the interstellar UV-field (Mamon et al.\ 1988). 
In the case of Mira, this value is relatively uncertain.
We apply an outer radius of the CO shell of 
2$\times 10^{16}\,\mbox{cm}$,  which corresponds to a 
diameter of about 20$\arcsec$ at the distance of Mira of 128\,pc, cf. Planesas et al. 
(1990b)\nocite{planesas:II}.  

\section{RESULTS PRODUCED BY THE MODEL}

For objects with low mass-loss rates, the NIR,
vibrational-rotational lines are sensitive to the stellar intensity, whereas 
pure rotational lines, such as the low rotational lines measured at radio 
wavelengths and the higher rotational lines observed with ISO 
are usually more sensitive to the temperature structure. 
The vibrational-rotational lines are therefore probes of the light-absorbing 
properties of the inner regions. 


The estimated fluxes given by the model are sensitive to many of the input 
parameters. We have run our model by fixing as many parameters as possible 
from the literature. Our variable parameters are the temperature ($T_0$) 
at a distance from the star of 10$^{15}$\,cm, the mass-loss rate 
($\dot{M}$), the terminal expansion velocity ($v_\mathrm{e}$) and turbulent 
velocity ($v_\mathrm{t}$), as well as the inner radius of the CSE 
($r_{\mathrm i}$).

Initially we apply a simple, `standard model' of a spherically symmetric 
and smooth CSE expanding at a constant velocity. 
Our observational constraints on the structure of the outer part of the wind 
are given mainly by the radio data, and
to a lesser extent, the far-IR ISO data in the same manner as by \citet{FL}. 
We later confront this model with the new observational constraints 
consisting of the absolute fluxes, the radial dependence of the scattered 
intensity as a function of angular distance, and the two line ratios
obtained from photospheric light scattered in vibrational-rotational lines of
circumstellar CO.

\subsection{The 'standard' model}

Previous radiative transfer models analysing the CSEs around bright M stars
(e.g., Kahane and Jura, 1994\nocite{kahane:94}) assume large velocity gradients (LVG). 
The model presented here is a more general code, based on the MC method, 
with a detailed radiative transport, including non-local effects, and treating 
the gas in full non-LTE. In the limit where the radiation field does not have 
a large influence [see for example the discussion on the `dusty' carbon star 
IRAS\,15\,194-5115 (Ryde et al. 1999c)\nocite{FL}], non-local effects are not 
very important. \citet{kahane:94} chose an inner radius of 
5$\times$10$^{14}$\,cm, i.e., approximately 13\,R$_*$, which is relatively far out 
in the wind. In our standard model we adopt this value.
Note that the dust is supposed to be situated well within this 
distance. 

The turbulent velocity and the expansion velocity of the gas in the CSE 
are free parameters in the model which we tune together with the 
other free parameters in order to explain the observed asymmetric radio-line profiles by 
self-absorption in an optically thick wind, i.e., the mass-loss rate has to be
sufficiently large. The temperature at
10$^{15}$\,cm (see Eq.\,\ref{T}) we set to
$T_0$$=$150\,K [a value consistent with those derived from a large survey of 
optically bright carbon stars (Sch\"{o}ier \& Olofsson 2000\nocite{Schoier}) 
where the kinetic temperature structure is obtained in a self-consistent 
manner]. 
The mass-loss rate is found to be 
2.5$\times$10$^{-7}$\,M$_{\sun}$\,yr$^{-1}$, the terminal expansion velocity 
is 2.5\,km\,s$^{-1}$, and the turbulent velocity 
required is 1.5\,km\,s$^{-1}$, see  Table\,\ref{parameters}.

In Figure\,\ref{model1} and \ref{model2} the model results are shown and compared with some
of the observations. 
We find that it is possible with this model to meet the 
observational constraints provided by the radio and ISO data reasonably well.
The three radio lines are reasonably well reproduced by the model
(Figure\,\ref{model1}a-c), suggesting a certain importance of self-absorption in the CSE around Mira. 
Also the modelled fluxes of the lines observed by ISO are reasonably well reproduced (Figure\,\ref{model1}d). 
The scattered, NIR fluxes, as a function of the distance 
from the star, are shown in Figure\,\ref{model2}a and 
they decline roughly as a power law, with a slope of $-3.4$ for the R(2) 
transition.
Figure\,\ref{model2}b shows the assumed kinetic-temperature law, 
as well as the excitation temperature  for some selected lines, 
throughout the envelope.
The excitation temperatures of the radio lines show that the populations of the lower lying 
levels of the scattered infrared lines R(1), R(2), and R(3) are not in local 
thermodynamic equilibrium (LTE).

In Figure\,\ref{model2}c the tangential optical depth 
is shown and gives an indication of the depths-of-formation of the different 
lines. 
Furthermore, in Fig.\,\ref{model2}d the relative population of level $i$ for 
the different rotational levels of the two lowest vibrational states of CO 
for our model are shown as a function of the distance from 
the star. 
The populations of the rotational states of the 
first excited vibrational state are several orders of magnitude 
lower than those of the ground level. 
Furthermore, it can be seen in the figure that these populations 
vary significantly with the distance from the star, a fact that can not
be taken into account in the analytic approach of \citet{ryde:apj}.

The angular behaviour of the vibrational-rotational line-intensities and
the line ratio of the two R-branch transitions are reasonably well reproduced
(Table\,\ref{resultat1}).
However, the intensities come out significantly too weak, approximately a 
factor of 5$-$10 lower than observed.
The maximum optical depth in the line-of-sight is approximately 25, while
the radial optical depth is approximately a factor of two higher.
We find that by varying the mass-loss rate, the temperature structure and
the turbulent velocity, it is not possible to model consistently all the 
observations. 

By introducing a velocity law, i.e., an accelerating wind, it is possible
to increase the NIR fluxes derived from the model.
Assuming a linearly increasing, with radius, expansion velocity of the CSE, 
NIR fluxes that are about 2 times larger than in the standard model is possible to
obtain.




\subsection{The `cavity' model}

We find that it is impossible to reach the high fluxes measured in the NIR 
lines far out in the CSE with a standard wind model extending all the 
way to the vicinity of the star. 
There are numerous ways of explaining an observed NIR flux that is lower 
than expected, but for a higher flux there are not many alternatives. 
The key problem is to prevent the innermost region of a smooth envelope to 
lose the radiation by scattering events in a geometrically diverging velocity 
field. In order to increase the NIR fluxes one can introduce an inner, 
empty region - i.e., postulate a large inner radius, which will increase the 
NIR flux at the point where the observations are made. 


By moving the inner radius of the CSE further out, high-density material close to the star is 
removed allowing the central radiation-field to penetrate further out into the wind. 
This will increase the excitation of the CO molecule, including the low-energy rotational
transitions. With a cavity it is possible to explain reasonably well the FIR fluxes 
as well as the observations of the NIR vibrational-rotational lines of CO, 
cf. Tables\,\,\ref{resultat1} and \ref{resultat2}. However, the radio-line intensities will 
come out to strong  in the model, due to the efficient radiative excitation, 
and the self-absorption present in these lines will become less pronounced since the optical depth 
decreases (by approximately a factor of 5 in the radial direction). 

We find that an inner boundary at 2.5$\times$10$^{15}$\,cm (approximately $1''$ or 
53 stellar radii) 
combined with a mass-loss rate of 
$\dot{M}$$=$2.5$\times$10$^{-7}$\,M$_{\sun}$\,yr$^{-1}$, i.e., the same as  
in `the standard model',
will result in a sufficiently high flux of stellar 
radiation reaching the wind regions beyond 2$\arcsec$ and at the same time 
enough scatterers at the location of our observations, i.e., at angular 
distances of 2$\arcsec$$-$7$\arcsec$ away from the star. 
The ISO lines are formed further in than the NIR vibrational-rotational 
lines, and will start to become sensitive to the adopted inner radius.
In the modelling, we have chosen the same 
temperature structure, velocity structure and turbulent velocity as was
obtained from `the standard model'.

The results of our best `cavity model' of the CSE around Mira are presented 
in Tables\,\ref{resultat1} and \ref{resultat2}, 
where the modelled intensities, line intensity ratios, intensity slopes, 
the modelled radio-fluxes, and the modelled mean ISO-flux
are compared with the observations. 

The modelled intensities of the three NIR lines 
[R(1), R(2), and R(3)] all decline in a $\log I - \log \beta$-plot with a 
slope of approximately $-$3, which is in good  agreement with the observations.
Note that there are systematic uncertainties in the  R(1) line due to the 
detector, which will lead to an uncertain R(1)/R(3) ratio [see the 
discussion by \citet{ryde:apj}; this has to be borne in mind when 
discussing the line ratios].



 



\section{DISCUSSION}

The Kitt Peak observations provide our modelling with new, strong constraints. 
These are the NIR absolute fluxes of the three lowest 
vibrational-rotational transitions in the R-branch of CO, their radial 
distributions, and the two line ratios from the three vibrational-rotational 
lines observed.

We find that it is not possible to meet these new constraints with a 
`standard model' of the wind, calculated on the basis of the several far-IR 
rotational transitions within the ground vibrational level of CO as observed
by ISO, and radio observations of millimeter rotational transitions of CO; 
our modelled NIR fluxes are an order of magnitude too low. 

By introducing a cavity 
in order to simulate  a higher flux 
at the point of scattering and keeping the mass-loss rate 
relatively high,  however, we are able to 
model the NIR fluxes and the brightness distribution, i.e., the 
decline in the intensity as a function of the distance from the star, providing 
a slope (d$\log$ I/d$\log \beta$) of approximately $-$3.
This results in few scatterers on the way and hence increase the number of 
photons that will be able to be scattered, and,
on the other hand, maintains enough scatterers at the point of our observations 
for the modelled fluxes  to be large enough. 
The important free parameters that couple to each other to give the number 
density of scatterers are the inner radius ($r_{\mathrm i}$), the mass-loss rate 
($\dot{M}$), and the expansion velocity ($v_\mathrm{e}$).

Thus, in order to reproduce the observed, high NIR fluxes, the radial optical
depth should not be too large. If there are too many scatterers on the way 
radially, many photons will scatter `sideways' due to the shifts out of 
the absorption profile caused by the velocity of the flow 
at a point close to the star and will never make it out to our point of observation. 
The high optical depths trap 
the photons as was suggested already by \citet{dyck}.
Thus, the velocity field is of great importance. 
If the {\it geometrical} velocity 
gradients are large across one optical depth unit 
(the mean distance between scattering 
events), the absorption profiles of 
the absorbing molecules will be shifted with respect to the emission profile 
of emitting molecules, leading to a decrease in the optical length for the 
emitted photons. This will lead to a loss of photons and a steepening of the 
decline of the intensity with distance from the star.

An important quantity for the flux levels at a certain point in the wind
is the relation between the expansion velocity, and the 
turbulent velocity (providing most of the width of the inherent line profiles).
If the turbulent velocity, which
is of the order of 1\,km\,s$^{-1}$, is comparable to the wind velocity, 
the velocity fields will be very important. 
Also in the case of Mira, this makes the modelling of the CSE sensitive to the
ratio between the turbulent velocity and the flow velocity. This is also what 
we find; a sensitive parameter in the modelling is this ratio. 

A physical reason for a higher flux could be a void of CO 
molecules, for instance due to an excess UV flux from the white-dwarf 
companion-star that certainly will partly dissociate CO,  or a
cavity caused by a decline of the stellar wind to a lower expansion velocity 
and a lower mass-loss rate, resembling the double wind scenario of 
\citet{knapp:98}. Other reasons could be that the gas close to the star is
in such a state that light is able to pass through relatively unaffected, 
either due to the medium being clumpy with a low filling factor, 
or by the matter being in radial 
structures in the innermost part of the CSE, which, 
further-out, developes into more smooth or shell-like structures. 
These structures could
occur depending on the expulsion mechanism of matter, such as a non-homogeneous mass-loss 
due to large convective eddies or 
magnetic spots, see for example \citet{soker:99}.
Further reasons could be CO
molecules absorbed onto dust grains, which are evaporated further out, or a 
small radial velocity gradient (i.e., the velocity slowly approaches the 
terminal velocity of the wind).


For large radial optical depths, the slope (d$\log$ I/d$\log \beta$)  will 
not be constant with distance as is suggested by the observations.
In models for which the optical depths are taken to be very large, we 
find that the slope of the R(3) line is the largest. 
As a consequence, the line ratios will vary greatly as a function of distance
from Mira which again is not observed. 
The line ratios R(1)/R(3) and R(2)/R(3) 
increase the further away from the star they are measured. 
(In a less optically thick model, the slopes will not differ greatly between the 
three lines and thus neither will the line ratios.)
In our observations, however, we will not be able to trace such a trend, having 
in mind the constraints due to signal-to-noise ratio and the
uncertainties involved.


The mass-loss rate determined here,
2.5$\times$10$^{-7}$\,M$_\odot$\,yr$^{-1}$, is in good
agreement with the mass-loss rate we estimated using an analytic approach to 
Mira's wind \citep{ryde:apj}. Note, however, that 
in this approach the level densities of CO are not known and a mean
is estimated, which will lead to an uncertainty, cf. Fig.\,\ref{pop}. 
Also, an assumption in the derivation of the analytical expression 
for the mass-loss rate is a {\it tangentially} optically thin wind, 
which is not strictly the case 
for a spherically symmetric and homogeneous wind. In a {\it tangentially} 
optically thin wind we expect a decline of the intensity with angular distance, given by a power law with
an exponent of $-3$. However, even in a {\it tangentially} optically thick wind we
are able to obtain the same d$\log$ I/d$\log \beta$$=$$-$3 slope.

The asymmetries in the radio lines, which are optically thick, are caused, 
in our `standard model', by self absorption along the line-of-sight.
Thus, we are able to reproduce the intensities and radio-line profiles 
reasonably well
without invoking a double-wind scenario as has previously been done in the literature
(cf. e.g., Knapp et al., 1998\nocite{knapp:98}).


\section{CONCLUSIONS}

In order to reproduce the intensity levels and the angular dependence of the 
intensity of the vibrational-rotational lines of the CO fundamental band 
at 4.6\,$\mu$m ($v$$=$1$\rightarrow$0), it is necessary to require low radial 
optical-depths to the point of scattering. In a standard CSE 
model, i.e., a smooth and spherically symmetric wind with an inner radius 
located fairly close to the central star, the radial optical 
depths are very large. 
The photons are scattered already close to the star and, due to the velocity 
field, shifted out of the absorption profile, leading to an escape.
The CSE model that is able to meet these new observational constraints 
reasonably well needs a form of inner cavity, devoid of gaseous CO.
The stellar light is then able to reach further out 
before it is scattered; the optical depths are reduced significantly.
The two CSE models presented here both require a mass-loss rate
of 2.5$\times$10$^{-7}$\,M$_{\sun}$\,yr$^{-1}$ and a turbulent velocity 
of 1.5\,km\,s$^{-1}$ given an expansion velocity of the wind 
of 2.5\,km\,s$^{-1}$. The only parameter that is changed between the 
two models is the location of the inner radius. In the standard model, that best
reproduces the observed radio and FIR emission, we use an inner radius located at approximately
13\,R$_*$ while for the  the cavity model, adopted to meet the NIR observational constraints,  
its location is approximately 50\,R$_*$.


We find that the ratio between the turbulent velocity and
the expansion velocity, is of great importance. 
The flux at 2$\arcsec$ is sensitive to this ratio;
the lower the ratio the lower the observed flux since more photons are lost
`sideways' due to velocity shifts out of the absorption profile. 
Furthermore, we conclude that for modelling vibrational-rotational lines the 
approximation of CRD is not a bad assumption as long as the optical depths are 
not too large.

The NIR observations in combination with a model of the CSE
is providing us with new pieces of information on the inner-most regions of
the wind.
The properties of this inner wind, that is simulated in the model by
a void of CO, can be explained physically by structures of 
the gas close to the star (within approximately 2$\arcsec$) that are in such 
a form that light is able to pass through. 
Further observations of the gas in the stellar wind of Mira slightly closer 
to the star could therefore be rewarding.


The inner parts of the circumstellar environment are complicated
regions of which we have a very limited understanding (see e.g.,  Ryde
et al., 1999a\nocite{ryde_CO2}), and where many of the basic assumptions of
our model are certainly not valid.
Thus, the model could fail in describing reality. We need further 
observational constraints.
The Phoenix spectrometer makes it possible to use long-slit spectroscopy, 
which makes spatially resolved spectra along the slit attainable. The slit is then 
used to map the CSE.
However, no interferometric mappings of the CSE of 
Mira with a high enough spatial resolution exist. Planesas et al. (1990b) presented  
observations with a 
resolution of 7.8$\arcsec$$\times$5.4$\arcsec$ using the Owens Valley millimeter-wave 
interferometer. The IRAM interferometer at 
Plateau de Bure can provide a spatial resolution of
under 1$\arcsec$. 

\begin{acknowledgements}

We are very greatful for inspiration from, and enlightening discussions 
with B. Gustafsson, H. Olofsson, and K. Eriksson. The referee is thanked
for valuable suggestions.
This research was supported by the Swedish National
Space Board, the Swedish Natural Science Research Council, 
and the Royal Swedish Academy of Sciences.

\end{acknowledgements}

\clearpage
\onecolumn

\figcaption[MIRA_light_curve.eps]{The {\it o}~Ceti light-curve.
1998 October 29 corresponds to a Julian date of 2451116.  At that time
{\it o}~Ceti had a visual magnitude of 7.8 and was in a pre-maximum
phase. (Mattei, J. A., 1999, Observations from the AAVSO International
Database, private communication). \label{fig3}}

\plotone{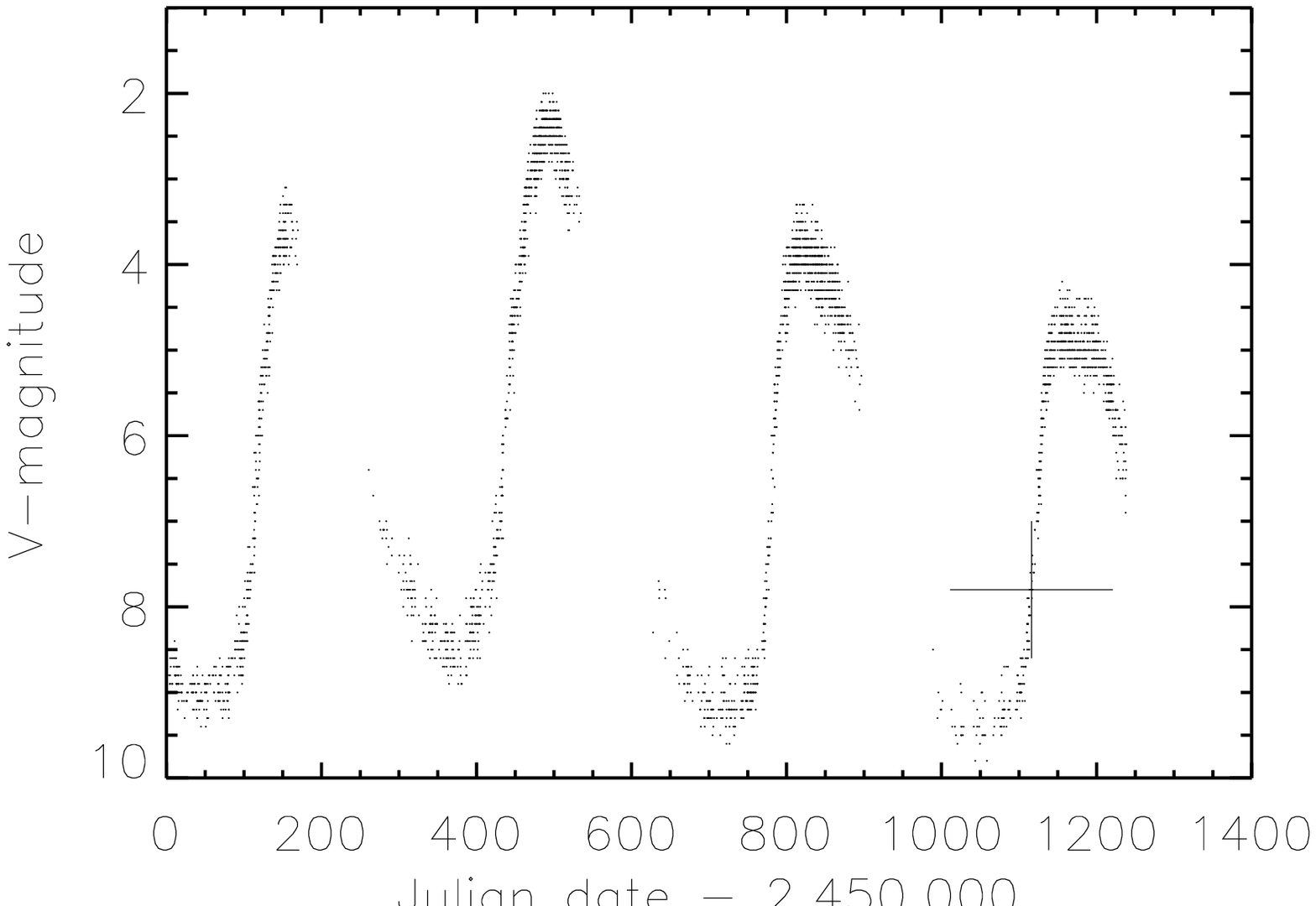}

\clearpage
\onecolumn

\figcaption[plot1.eps]{Modelling results for the best standard model of Mira's 
wind. a-c) The JCMT millimeter-wave observations, in main-beam brightness temperature, 
overlayed with the model line-profiles. The beam size used is indicated
in the upper right-hand corner of each panel. Note that for the CO($J$$=$$3$$\rightarrow$$2$) we 
also show a scaled line
profile for easier comparison with the observed shape of the emission line.
d) ISO observations and the model overlays using a beam size of 70$\arcsec$.\label{model1}}

\plotone{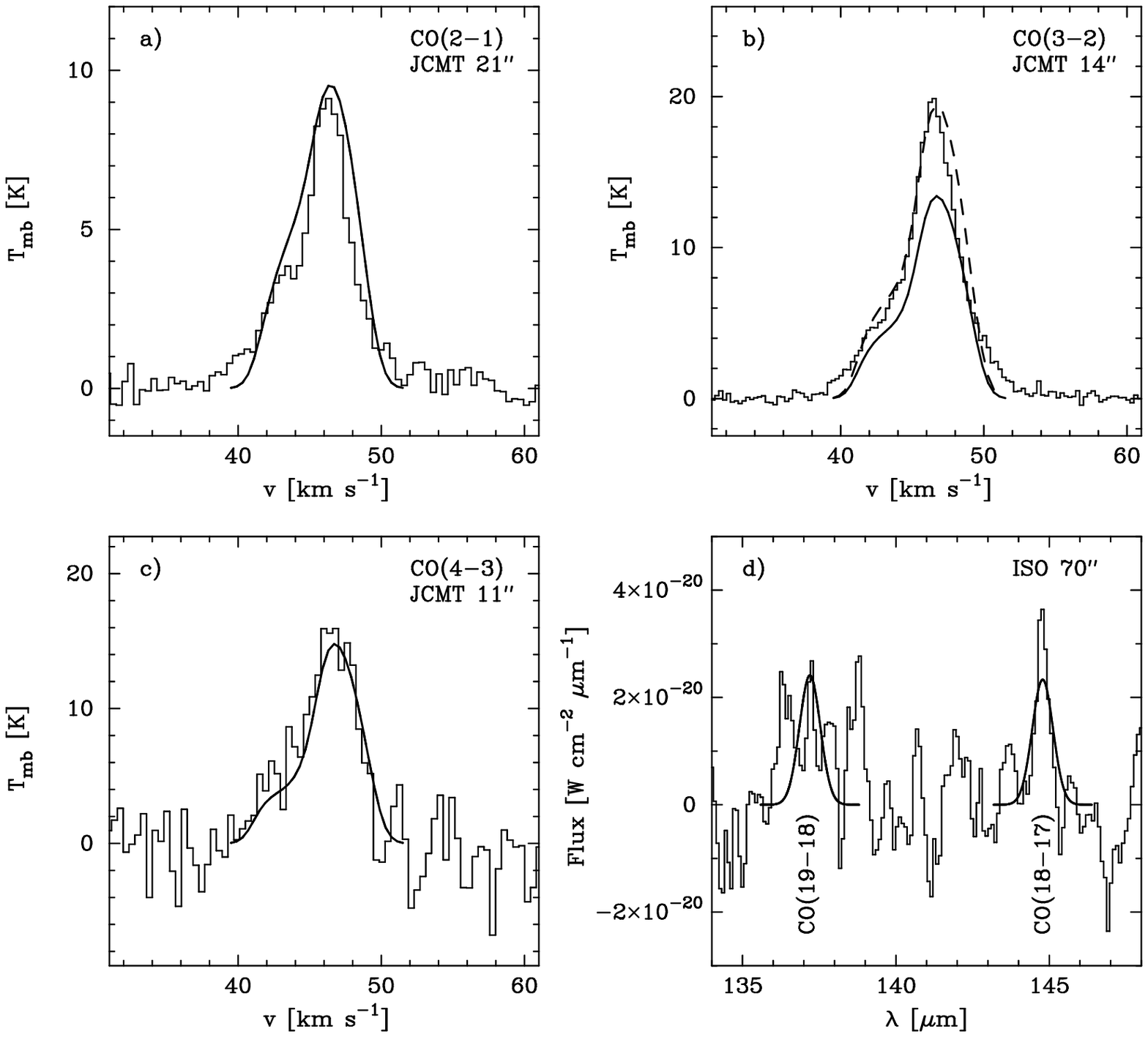}

\clearpage
\onecolumn

\figcaption[plot2.eps]{Modelling results for the best standard model of Mira's 
wind. a) The intensity of the NIR emission lines R(1) (full), 
R(2) (dashed), and R(3) (dash-dotted) 
as a function of the distance (in cm and arcseconds) from the star. 
b) The kinetic temperature structure of the model of the 
circumstellar envelope, given by a power law,
is shown as a thick full line. 
The three other lines show the
calculated excitation-temperature structure of the infrared R(2)-line 
(dotted), the radio CO($J$$=$$3$$\rightarrow$$2$) line (dashed), 
and the far-infrared CO($J$$=$$18$$\rightarrow$$17$) line (dash-dotted).
All lines are non-thermal.
The CO($J$$=$$18$$\rightarrow$$17$) line is shown only out to 
$r$$=$1$\times$10$^{16}$\,cm since beyond this radius these levels are scarcely populated, 
see Fig.\,\ref{pop}$^{\mathrm d}$.
c) The tangential optical depths of three selected lines (symbols same as in b). 
These give an indication of how far out in the wind the lines are formed. 
d) Relative level-populations 
according to our best model of Mira's wind. 
In the model we include 30 rotational levels each in the two 
lowest-lying vibrational states. For clarity, only 10 rotational levels from each of the two
vibrational levels are shown.  The upper set of graphs represent the populations of the rotational levels
in the ground vibrational state ($v$$=$$0$), i.e., $J$$=$$0$$-$$10$. 
The lower set of graphs represent the 
excited vibrational state $v$$=$$1$. For instance, the R(1) line is a transition from ($v$$=$$1$, $J$$=$$2$) to
($v$$=$$0$, $J$$=$$1$). 
\label{pop}
\label{model2}}

\plotone{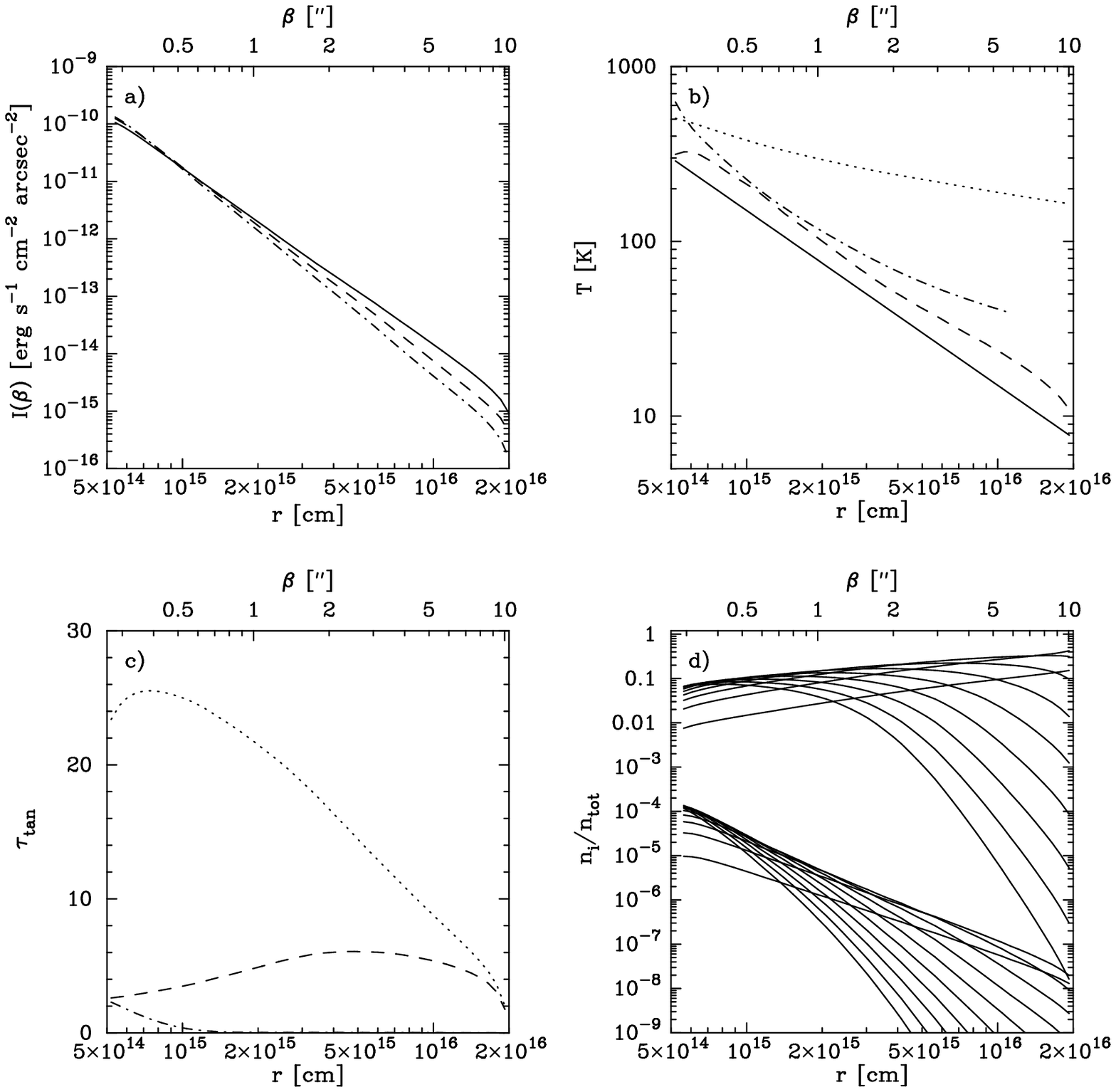}

\clearpage

\begin{deluxetable}{ll}
\tablewidth{0pt}
\tablecaption{The parameters adopted for the CSE around {\it o} Ceti \label{parameters}}
\tablehead{  
 \colhead{Model parameter} & \colhead{Value}   \\
}
\startdata
 $T_*$ &  2400\,K  \\
  \noalign{\smallskip} 
 $T_\mathrm{d}$ & 1100\,K\\
  \noalign{\smallskip}
 $L_\mathrm{bol}$ &  8900\,L$_\odot$\\
  \noalign{\smallskip}
 $L_*$ & 7100\,L$_\odot$\\
  \noalign{\smallskip}
 $L_\mathrm{d}$ & 1800\,L$_\odot$  \\
  \noalign{\smallskip}
 $R_*$  & 3.8$\times$10$^{13}$\,cm \\
  \noalign{\smallskip} 
 $R_\mathrm{d}$ & 3\,R$_*$ \\
  \noalign{\smallskip}
 $D$    & 128\,pc \\
  \noalign{\smallskip}
 $\dot{M}$ &  2.5$\times$10$^{-7}$\,M$_\odot$\,yr$^{-1}$ \\
  \noalign{\smallskip}
 $f_\mathrm{CO}$ & 5$\times$10$^{-4}$\\
  \noalign{\smallskip}    
 $v_{\mathrm e}$                  &  2.5\,km\,s$^{-1}$  \\
  \noalign{\smallskip}
 $r_{\mathrm p}$  &  2.0$\times$10$^{16}$\,cm\\
  \noalign{\smallskip}
 $r_{\mathrm{i, standard}}$  &  5.0$\times$10$^{14}$\,cm \\
   \noalign{\smallskip}
 $r_{\mathrm{i, cavity}}$  &  2.5$\times$10$^{15}$\,cm \\
   \noalign{\smallskip}
 $v_\mathrm{t}$ & 1.5\,km\,s$^{-1}$  \\ 
  \noalign{\smallskip} 
 $T_0$  & 150\,K \\
  \noalign{\smallskip} 
 $R_0$  &  1.0$\times$10$^{15}$\,cm\\

\enddata

\end{deluxetable}

\clearpage

\begin{deluxetable}{l l l l l l l}
\tablewidth{0pt}

\tablecaption{Intensities in 10$^{-12}$\,erg\,s$^{-1}$cm$^{-2}$arcsec$^{-2}$ (measured at $2\arcsec$)
, the line intensity ratios (averaged over $2-3.4\arcsec$),
 and the variation of the R(2) intensity with distance from the star.
\label{resultat1}}
\tablehead{  
\colhead{}    &\colhead{$I$[R(1)]} & \colhead{$I$[R(2)]} & \colhead{$I$[R(3)]} 
& \colhead{$\frac{\mathrm{R(1)}}{\mathrm{R(3)}}$} & \colhead{$\frac{\mathrm{R(2)}}{\mathrm{R(3)}}$} 
& \colhead{$\frac{\mathrm{d}\log \mathrm{I[R(2)]}}{\mathrm{d}\log \beta}$}}
\startdata
Observed: &     2.1$\pm$1.0 & 1.5$\pm$0.8 & 1.2$\pm$0.6  & 1.7$\pm$0.6  & 1.2$\pm0.2$ & -2.8$\pm$0.4 \\ 
 Modelled (`standard'): &     $0.24$ & $0.16$ & $0.11$ &  $2.2$  & $1.5$  & -3.4 \\ 
 Modelled (`cavity'):   &     $1.21$  & $1.18$ & $1.12$ &  $1.3$  & $1.1$  & -3.0 \\ 
\enddata

\end{deluxetable}

\clearpage

\begin{deluxetable}{l l l l l}
\tablewidth{0pt}

\tablecaption{The integrated radio flux
in K\,km\,s$^{-1}$, and the averaged ISO flux in 10$^{-20}$\,W\,cm$^{-2}$. 
\label{resultat2}}   
\tablehead{  
\colhead{} &  \colhead{CO($J$$=$2$\rightarrow$1)} &  \colhead{CO($J$$=$3$\rightarrow$2)} & 
\colhead{CO($J$$=$4$\rightarrow$3)} & \colhead{ISO flux} }

\startdata 
Observed: & 44$\pm$9 & 94$\pm$19 & 77$\pm$15 & 2$\pm$1\\ 
 Modelled (`standard'):  & 50 & 67  & 71  & 2.0 \\ 
 Modelled (`cavity'):    & 72 & 147 & 178 & 2.1 \\ 
\enddata
\end{deluxetable}


\begin{thebibliography}{}

\bibitem[\protect\astroncite{{Bernes}}{1979}]{bernes}
{Bernes}, C., 1979,
\newblock {A\&A} {73}, 67

\bibitem[\protect\astroncite{{Bester} et~al.}{1991}]{bester:91}
{Bester}, M., {Danchi}, W.~C., {Degiacomi}, C.~G., {Townes}, C.~H., and
  {Geballe}, T.~R., 1991,
\newblock {ApJ} {367}, L27

\bibitem[\protect\astroncite{{Clegg} et~al.}{1996}]{LWS}
{Clegg}, P., {Ade}, P., {Armand}, C., et~al., 1996,
\newblock {A\&A} {315}, L38

\bibitem[\protect\astroncite{{Crosas} and {Menten}}{1997}]{crosas97}
{Crosas}, M. and {Menten}, K.~M., 1997,
\newblock {ApJ} {483}, 913

\bibitem[\protect\astroncite{{Crosas} et~al.}{1997}]{crosas:apss}
{Crosas}, M., {Menten}, K.~M., {Young}, K., and {Phillips}, T.~G., 1997,
\newblock {Ap\&SS} {251}, 189

\bibitem[\protect\astroncite{{Danchi} et~al.}{1994}]{danchi:94}
{Danchi}, W.~C., {Bester}, M., {Degiacomi}, C.~G., {Greenhill}, L.~J., and
  {Townes}, C.~H., 1994,
\newblock {AJ} {107}, 1469

\bibitem[\protect\astroncite{{Dyck} et~al.}{1983}]{dyck}
{Dyck}, H.~M., {Beckwith}, S., and {Zuckerman}, B., 1983
\newblock {ApJ} {271}, L79

\bibitem[\protect\astroncite{ESA}{1997}]{hipp}
ESA, 1997,
\newblock ESA,
\newblock The Hipparcos and Tycho Catalogues, ESA SP-1200

\bibitem[\protect\astroncite{{Glassgold}}{1996}]{glassgold}
{Glassgold}, A.~E., 1996,
\newblock {ARA\&A} {34}, 241 

\bibitem[\protect\astroncite{{Goldreich} and {Scoville}}{1976}]{goldreich76}
{Goldreich}, P. and {Scoville}, N., 1976,
\newblock {ApJ} {205}, 144

\bibitem[\protect\astroncite{{Groenewegen}}{1994}]{groenewegen94}
{Groenewegen}, M. A.~T., 1994,
\newblock {A\&A} {290}, 531


\bibitem[\protect\astroncite{{Gustafsson} and {Ryde}}{2000}]{IAU177}
{Gustafsson}, B. and {Ryde}, N., 2000,
\newblock {in: B. Wing (ed.) Proc. IAU Symp. 
177, The Carbon Star Phenomenon. Kluwer, Dordrecht, p. 481}


\bibitem[\protect\astroncite{{Haniff} et~al.}{1992}]{haniff:92}
{Haniff}, C.~A., {Ghez}, A.~M., {Gorham}, P.~W., et~al., 1992,
\newblock {AJ} {103}, 1662

\bibitem[\protect\astroncite{{Haniff} et~al.}{1995}]{haniff:95}
{Haniff}, C.~A., {Scholz}, M., and {Tuthill}, P.~G., 1995,
\newblock {MNRAS} {276}, 640

\bibitem[\protect\astroncite{{Hur\'e} and {Roueff}}{1996}]{hure}
{Hur\'e}, J.~M. and {Roueff}, E., 1996,
\newblock {A\&AS} {117}, 561

\bibitem[\protect\astroncite{{Kahane} and {Jura}}{1994}]{kahane:94}
{Kahane}, C. and {Jura}, M., 1994,
\newblock {A\&A} {290}, 183

\bibitem[\protect\astroncite{{Karovska} et~al.}{1997}]{karovska:97}
{Karovska}, M., {Hack}, W., {Raymond}, J., and {Guinan}, E., 1997,
\newblock {ApJ} {482}, L175

\bibitem[\protect\astroncite{{Karovska} et~al.}{1993}]{karovska:93}
{Karovska}, M., {Nisenson}, P., and {Beletic}, J., 1993,
\newblock {ApJ} {402}, 311

\bibitem[\protect\astroncite{{Karovska} et~al.}{1991}]{karovska:91}
{Karovska}, M., {Nisenson}, P., {Papaliolios}, C., and {Boyle}, R.~P., 1991,
\newblock {ApJ} {374}, L51

\bibitem[\protect\astroncite{{Kessler} et~al.}{1996}]{kessler}
{Kessler}, M.~F., {Steinz}, J.~A., {Anderegg}, M.~E., et~al., 1996,
\newblock {A\&A} {315}, L27

\bibitem[\protect\astroncite{{Kirby-Docken} and {Liu}}{1978}]{kirby}
{Kirby-Docken}, K. and {Liu}, B., 1978,
\newblock {ApJS} {36}, 359

\bibitem[\protect\astroncite{{Knapp} et~al.}{1998}]{knapp:98}
{Knapp}, G.~R., {Young}, K., {Lee}, E., and {Jorissen}, A., 1998,
\newblock {ApJS} {117}, 209

\bibitem[\protect\astroncite{{Kwan} and {Webster}}{1993}]{kwan}
{Kwan}, J. and {Webster}, Z., 1993,
\newblock {ApJ} {419}, 674

\bibitem[\protect\astroncite{{Lopez} et~al.}{1997}]{lopez:97}
{Lopez}, B., {Danchi}, W.~C., {Bester}, M., et~al., 1997,
\newblock {ApJ} {488}, 807

\bibitem[\protect\astroncite{{Mahler} et~al.}{1997}]{mahler:97}
{Mahler}, T.~A., {Wasatonic}, R., and {Guinan}, E.~F., 1997,
\newblock {Informational Bulletin on Variable Stars} {4500}, 1

\bibitem[\protect\astroncite{{Mamon} et~al.}{1988}]{mamon}
{Mamon}, G.~A., {Glassgold}, A.~E., and {Huggins}, P.~J., 1988,
\newblock {ApJ} {328}, 797

\bibitem[\protect\astroncite{{Netzer} and {Knapp}}{1987}]{netzer}
{Netzer}, N. and {Knapp}, G.~R., 1987,
\newblock {ApJ} {323}, 734

\bibitem[\protect\astroncite{{Neufeld} and {Kaufman}}{1993}]{neufeld}
{Neufeld}, D.~A. and {Kaufman}, M.~J., 1993,
\newblock {ApJ} {418}, 263

\bibitem[\protect\astroncite{{Olofsson}}{1996}]{ho:rev}
{Olofsson}, H., 1996,
\newblock {Ap\&SS} {245}, 169

\bibitem[\protect\astroncite{{Olofsson} et~al.}{1996}]{ho:96a}
{Olofsson}, H., {Bergman}, P., {Eriksson}, K., and {Gustafsson}, B., 1996,
\newblock {A\&A} {311}, 587

\bibitem[\protect\astroncite{{Planesas} et~al.}{990a}]{planesas:I}
{Planesas}, P., {Bachiller}, R., {Martin-Pintado}, J., and {Bujarrabal}, V.,
  1990a,
\newblock {ApJ} {351}, 263

\bibitem[\protect\astroncite{{Planesas} et~al.}{990b}]{planesas:II}
{Planesas}, P., {Kenney}, J. D.~P., and {Bachiller}, R., 1990b,
\newblock {ApJ} {364}, L9

\bibitem[\protect\astroncite{{Quirrenbach} et~al.}{1992}]{quirren:92}
{Quirrenbach}, A., {Mozurkewich}, D., {Armstrong}, J.~T., et~al., 1992,
\newblock {A\&A} {259}, L19

\bibitem[\protect\astroncite{{Ryde} et~al.}{1999a}]{ryde_CO2}
{Ryde}, N., {Eriksson}, K., and {Gustafsson}, B., 1999a,
\newblock {A\&A} {341}, 579

\bibitem[\protect\astroncite{{Ryde} et~al.}{2000}]{ryde:apj}
{Ryde}, N., {Gustafsson}, B., {Eriksson}, K., and {Hinkle}, K.~H., 2000,
\newblock {ApJ} in press


\bibitem[\protect\astroncite{{Ryde} et~al.}{1999b}]{ryde:letter}
{Ryde}, N., {Gustafsson}, B., {Hinkle}, K.~H., {Eriksson}, K., {Lambert},
  D.~L., and {Olofsson}, H., 1999b,
\newblock {A\&A} {347}, L35

\bibitem[\protect\astroncite{{Ryde} et~al.}{1999c}]{FL}
{Ryde}, N., {Sch\"oier}, F.~L., and {Olofsson}, H., 1999c,
\newblock {A\&A} {345}, 841

 
\bibitem[\protect\astroncite{{Sch\"oier}}{2000}]{PhD}
{Sch\"oier}, F.~L., 2000,
\newblock PhD thesis, Stockholm Observatory

\bibitem[\protect\astroncite{{Sch\"oier}}{2000}]{Schoier}
{Sch\"oier}, F.~L., and {Olofsson}, H., 2000,
\newblock  A\&A submitted


\bibitem[\protect\astroncite{{Soker} and {Clayton}}{1999}]{soker:99}
{Soker}, N.  and {Clayton}, G.~C., 1999,
\newblock {NMRAS} {307}, 993

\bibitem[\protect\astroncite{{Stickland} et~al.}{1982}]{stickland}
{Stickland}, D.~J., {Cassatella}, D., and {Ponz}, D., 1982,
\newblock {MNRAS} {199}, 1113

\bibitem[\protect\astroncite{{Stanek} et~al.}{1995}]{stanek:95}
{Stanek}, K.~Z., {Knapp}, G.~R., {Young}, K., and {Phillips}, T.~G., 1995,
\newblock {ApJS} {100}, 169

\citet{stickland}

\bibitem[\protect\astroncite{{Swinyard} et~al.}{1996}]{swinyard}
{Swinyard}, B., {Clegg}, P., {Ade}, P., et~al., 1996,
\newblock {A\&A} {315}, L43

\bibitem[\protect\astroncite{{Van Leeuwen} et~al.}{1997}]{leeuwen:97}
{Van Leeuwen}, F., {Feast}, M.~W., {Whitelock}, P.~A., and {Yudin}, B., 1997,
\newblock {MNRAS} {287}, 955

\bibitem[\protect\astroncite{{Wilson} et~al.}{1992}]{wilson:92}
{Wilson}, R.~W., {Baldwin}, J.~E., {Buscher}, D.~F., and {Warner}, P.~J., 1992,
\newblock {MNRAS} {257}, 369

\bibitem[\protect\astroncite{{Young}}{1995}]{young}
{Young}, K., 1995,
\newblock {ApJ} {445}, 872


\end{thebibliography}
\end{document}